# Static Malware Detection & Subterfuge: Quantifying the Robustness of Machine Learning and Current Anti-Virus

William Fleshman[1,4], Edward Raff[1,2,3] Richard Zak[1,2], Mark McLean[1], and Charles Nicholas[3]

[1] Laboratory for Physical Sciences
{william.fleshman, edraff, rzak, mrmclea}@lps.umd.edu
[2] Booz Allen Hamilton
{raff_edward, zak_richard}@bah.com
[3] University of Maryland, Baltimore County
{raff.edward, nicholas}@umbc.edu
[4] U.S. Army
william.c.fleshman.mil@mail.mil

**Abstract.** As machine-learning (ML) based systems for malware detection become more prevalent, it becomes necessary to quantify the benefits compared to the more traditional anti-virus (AV) systems widely used today. It is not practical to build an agreed upon test set to benchmark malware detection systems on pure classification performance. Instead we tackle the problem by creating a new testing methodology, where we evaluate the change in performance on a set of known benign & malicious files as adversarial modifications are performed. The change in performance combined with the evasion techniques then quantifies a system's robustness against that approach. Through these experiments we are able to show in a quantifiable way how purely ML based systems can be more robust than AV products at detecting malware that attempts evasion through modification, but may be slower to adapt in the face of significantly novel attacks.

## 1 Introduction

The threat and impact of malicious software (malware) has continued to grow every year. The annual financial impact is already measured in the hundreds of billions of dollars [2, 11]. Simultaneously, there are worries that the classical anti-virus approach may not continue to scale and fail to recognize new threats [36].

Anti-virus systems were historically built around signature based approaches. Wressnegger et al. [38] discussed a number of issues with signatures, but the primary shortcoming is an intrinsically static nature and inability to generalize. Most anti-virus companies have likely incorporated machine learning into their software, but to what extent remains unclear due to the nature of proprietary information. Regardless, adversaries can still successfully evade current detection systems with minimal effort. The success of methods like obfuscation and polymorphism is evident by its prevalence, and recent work has suggested that the majority of unique malicious files are due to the use of polymorphic malware [19].

Given these issues and urgency, Machine Learning would appear to be a potential solution to the problem of detecting new malware. Malware detection can be directly phrased as a classification problem. Given a binary, we define some feature set, and learn a binary classifier that outputs either benign or malicious. The output of this model could be calibrated to reach any desired false-positive ratio.

However, one should never switch to a new technology for its own sake. It is necessary to have empirical and quantifiable reasons to adopt a new approach for malware detection. Ideally, this would be based off the accuracy of current anti-virus systems compared to their machine-learning counterparts. In reality, this is non-trivial to estimate, and many arguments currently exist as to how this should be done [6, 13, 35]. Designing a corpus and test protocol to determine accuracy at large is hampered by issues like concept drift [12], label uncertainty, cost [20], and correlations over time [14, 34].

Toward remedying this issue, we propose a new testing methodology that can highlight a detector's strength or weakness with regard to specific attack types. We first utilize the framework from Biggio et al., to specify a model for our adversary [4]. After defining the adversary's goal, knowledge, and capabilities, we look at how difficult it is for an adversary with known malware to evade malware detection systems under comparison. As examples, we compare two machine learning approaches, one a simpler n-gram model and one neural network based, to a quartet of production AV systems. While we do not exhaustively test all possible evasion techniques, we show the potential for our protocol using a handful of relevant tests: non-destructive automated binary modification, destructive adversarial attacks, obfuscation through packing, and malicious injection. We are specifically looking only at black box attacks which can be applied to any potential malware detector, assuming that the adversary is looking for maximal dispersion and not a targeted attack. This new evaluation protocol is our primary contribution, and a black-box adversarial attack for byte-based malware detectors our second contribution. This evaluation protocol allows us to make new conclusions on the potential benefits and weakness of two byte based machine learning malware detectors.

We will review previous work in section 2. The experiments and methodology we use to compare our ML based systems to anti-virus will be given in section 3 and section 4, followed by their results in section 5. We will discuss the higher level takeaways in section 6, and then conclude in section 7.

## 2  Related Work

Malware detection and analysis has been an area of active research for several years. One class of techniques that is of particular note is *dynamic analysis*, where the binary itself is run to observe its behavior. Intuitively, malicious behavior is the best indicator of a malicious binary — making dynamic analysis a popular approach [8]. However, dynamic analysis has many complications. It requires significant effort and infrastructure to make it accurately reflect user environments, which are often not met in practice [29]. Furthermore, the malware author can use a number of techniques to detect dynamic execution, hide behavior, or otherwise obfuscate such analysis. This means effective dynamic analysis systems are often a cat-and-mouse game of technical issues, and can require running binaries across many layers of emulation [7].

It is for these reasons that we focus on the static analysis case. This removes the unlimited degrees of freedom provided by actually running code. However, this does not mean the malicious author has no recourse. Below we will review the related work in this area.

Questions regarding the security of machine learning based systems have been an active area of research for over a decade [3, 5], but have recently received increased attention. This is in particular due to the success in generating *adversarial* inputs to neural networks, which are inputs that induce an incorrect classification despite only minor changes to the input [9, 37]. These approaches generally work by taking the gradient of the network with respect to the input, and adjusting the input until the network's output is altered. This does not directly translate to malware detection when using the raw bytes, as bytes are discontinuous in nature. That is to say, any change in a single byte's value is an equally "large" change, whereas in images, adjusting a pixel value can result in visually imperceptible changes. Arbitrary byte changes may also result in a binary that does not execute, making it more difficult to apply these attacks in this space [10, 30]. While such attacks on individual models are possible [15, 17], we are concerned with attacks that can be applied to any detector, therefore these methods are out of the scope of this work.

To overcome the issue with arbitrary byte changes breaking executables, Anderson et al. [1] developed a set of benign actions that can modify a binary without changing its functionality. Selecting these actions at random allowed malware to evade a ML model 13% of the time. Introducing a Reinforcement Learning agent to select the actions increased this to 16% of the time. Their ML model used features from the PE header as well as statistics on strings and bytes from the whole binary. We will replicate this type of attack as one of the comparisons between our ML systems and anti-virus.

Another approach is to modify the bytes of a binary, and run the malware detection system after modification. While this risks breaking the binary, it can still be effective, and it is important to still quarantine malware even if it does not execute as intended. Wressnegger et al. [38] modified each binary one byte at a time, and found that single byte changes could evade classical AV detectors. They used this approach to find which bytes were important to the classification, and reverse-engineered signatures used from multiple AV products. We extend this technique in subsection 4.2 to create a new adversarial approach against our models, which finds a contiguous byte region which can be altered to evade detection.

For the machine learning based malware detection methods, we look at two approaches: byte n-grams and neural networks. The byte n-gram approach was one of the first machine learning methods proposed for malware detection [32], and has received significant attention [16]. We use the same approach for training such models as presented in Raff et al. [28]. For the neural network approach we use the recently proposed MalConv, which processes an entire binary by using an embedding layer to map bytes to feature vectors followed by a convolutional network [26]. Because the embedding layer used by MalConv is non-differentiable with respect to the input bytes, the aforementioned adversarial approaches applied to image classification tasks are not as easily applicable. Similarly, anti-virus products do not provide a derivative for such attacks. This is why we develop a new black-box adversarial approach in subsection 4.2.

Using these modeling methods is beneficial for our analysis because they are trained with minimal assumptions about the underlying data. Other approaches to malware detection that process the PE-header [33] or use disassembly [21] exist. We avoid these under the belief that their use of explicit feature sets aids in evading them (e.g, modifying the PE-Header is easy and allows for direct evasion of models using those features).

We leverage the existing framework from [4] to model our adversary for each comparison made. This framework prescribes defining the adversary's goals, knowledge, and capabilities in regards to the classifiers.

## 3  Adversarial Model

For each experiment we specify the model of our adversary by outlining the goals, knowledge, and capabilities available [4]. These make clear the assumptions and use cases presented by each scenario.

The goal of our adversary is the same across all experiments – using a single attack to maximize the misclassification rate of malicious files across many classifiers. The one caveat is that in subsection 4.2 we only goes as far as identifying a small portion of the file which can be altered to evade detection. A real attacker would then have to alter those bytes while retaining the malicious functionality of the file. It is unlikely that authors of benign software would seek a malicious classification – therefore we limit ourselves to experiments on malware only.

We severely constrain the knowledge of our adversary. In all cases we assume the adversary has no knowledge of the classifiers' algorithms, parameters, training data, features, or decision boundaries. This is the best case scenario for security applications, as knowledge of any of the previous attributes would increase the sophistication of possible attacks.

Similarly, the experiments outlined in subsection 4.1, subsection 4.3, and subsection 4.4 require no prior interaction with the classifiers before deploying a malicious file to the wild. In subsection 4.2, the adversary has the capability of querying our n-gram model only. The inner workings of the model remain hidden, but the output of the model is available for deductive reasoning.

These assumptions are restrictive to the adversary's actions and knowledge, but are important because current adversaries are able to succeed under such restrictions today. This means malware authors can obtain success with minimal effort or expense. Through our evaluation procedure we are able to better understand which techniques / products are most robust to this restricted scenario, and thus will increase the effort that must be expended by malware authors.

## 4 Experiments and Methodology

We now perform experiments under our proposed testing methodology to compare machine learning models to commercial anti-virus systems. To do so, we compare two machine learning classifiers and four commercial anti-virus products: AV1, AV2, AV3, and AV4. We use these anonymous names to be compliant with their End User License Agreements (EULA), which bars us from naming the products in any comparative publication. We are prohibited from using the internet based mass scanners, like VirusTotal, for similar reasons. Ideally, we could test many products, but with our large corpus of files we limit ourselves to these four – which we believe to be representative of the spectrum of commercially used AV platforms. We purposely did not chose any security products which advertise themselves as being primarily powered by machine learning or artificial intelligence. Additionally, any cloud based features which could upload our files were disabled to protect the proprietary nature of our dataset. Our machine learning based approaches will be built upon n-grams and neural networks which process a file's raw bytes. We will compare these systems on a number of tasks to evaluate their robustness to evasion.

We now describe our new testing protocol. Given a corpus $\{x_1, \ldots, x_n\}$ of $n$ files with known labels $y_i \in \{\text{Benign}, \text{Malicious}\}$, and detector $C(\cdot)$ we first measure the accuracy of $C(x_i), \forall i$. This gives us our baseline performance. We then use a suite of evasion techniques $\phi_1(\cdot), \ldots, \phi_K(\cdot)$ and look at the difference in accuracy between $C(x_i) = y_i$ and $C(\phi(x_i)) = y_i$. The difference in these scores tells us the ability of the detector $C(\cdot)$ to generalize past its known data and catch evasive modifications. Any approach for which $C(x_i) \neq C(\phi_j(x_i))$ for many different evasion methods $\phi$ which preserve the label $y_i$, is then an intrinsically brittle detector, even if it obtains perfect accuracy before $\phi$ is used.

The raw detection accuracy is not the point of this test. Each system will have been built using different and distinct corpora of benign and malicious files. For the AV products, these are used to ensure that a certain target level of false positives and false negatives are met. In addition, the training corpus we use for our models is orders of magnitude smaller than what the AV companies will have access to. For this reason we would not necessarily expect our approaches to have better accuracy. The goal is to quantify the robustness of any classifier to a specific threat model and attack.

One novel and three pre-existing techniques ("$\phi$"s) will be used to perform this evaluation. These are not exhaustive, but show how our approach can be used to quantify relative performance differences. Below we will review their methodology, and the information that we can glean from their results. In section 5 we will review the results of running these experiments.

**Dataset**

We take only a brief moment to expound upon the machine learning models and data used for training, as the information disparity with other AV products makes direct comparison difficult. We trained the machine learning models on a large dataset from industry discussed in [27] which contains approximately 2 million samples balanced almost evenly between benign and malicious. Our test set consists of approximately 80,000 files held out for our post-training experimentation. The test set is also almost perfectly balanced. This set was not seen by our machine learning models during training, but could have files previously seen by the anti-virus corporations. The n-gram model follows the procedure used in [28], but uses one million features instead of 100,000. The MalConv approach is specified in [26]. Both of these approaches require no domain knowledge to apply.

### 4.1 Randomly Choosing Benign Modifications

In this experiment, we use a subset[5] of the modifications used by Anderson et al. [1] to alter malicious files before scanning them with the classifiers. The objective is to test the brittleness of the classifiers' decision mechanisms by making small changes that do not alter the functionality of the malware. This experiment is ideal because it produces a new file that is still functional, and should have no impact on an *ideal* malware detector.

There are nine different modification actions, and grouped by type are:

- rename or create new sections
- append bytes to the end of a section or the file
- add an unused function to the import table
- create a new entry point (which jumps to the old entry)
- modify the header checksum, the signature, or debug info

Each malicious file was scanned before the experiment to see if it was already being misclassified as benign (false negative). If it was correctly classified as malicious then a modification was randomly selected from the set and applied. This process was repeated up to ten times or until the file evaded the classifier. Approaches that rely too heavily on exact signatures or patterns should suffer in this test, and we would expect any approach that uses purely dynamic features to be unaffected by the modifications.

This experiment is time intensive, and so we use a subset of 1,000 files randomly chosen from our test set. Anderson et al. limited themselves to a small 200 file sample, so our larger test set may provide additional confidence in the results. The original paper proposed using a reinforcement learning algorithm (RLA) to select the modifications in an adversarial manner. We found that including a RLA did not change results, but did limit our ability to test with all products in a timely manner.

---

[5] A UPX (un)packing option did not operate correctly and so was excluded.

## 4.2 Targeted Occlusion of Important Bytes

The first approach described requires extensive domain knowledge based tooling, and would be difficult to adapt to new file types. We introduce a novel approach to find important byte regions of a malicious file given a working detector. Identifying the most important region of a file gives feedback to adversaries, and allows us to occlude that region as an evasion technique.

Our approach works by occluding certain bytes of a file, and then running the malware detector on the occluded file. By finding a contiguous region of bytes that reduced the detector's confidence that a file is malicious, we can infer that the bytes in that region are important to the detector. If occluding a small region causes the detector to change its vote from malicious to benign, then we can infer that the detector is too fragile. In this case it is plausible for a malware author to determine what is stored in the small region detected, and then modify the binary to evade detection.

Manually editing groups of bytes one at a time would be computationally intractable. Given a file $F$ of $|F|$ bytes, and a desired contiguous region size of $\beta$, it would take $O(|F|)$ calls to the detector to find the most important region. We instead develop a binary-search style procedure, which is outlined in Algorithm 1. Here $C(\cdot)$ returns the malware detector's confidence that a given file is malicious, and $\mathcal{D}$ is a source of bytes to use for the occlusion of the original bytes. This approach allows us to find an approximate region of importance in only $O(\log |F|)$ calls to the detector $C(\cdot)$. In the event that $C(F_l) = C(F_r)$, ties can be broken arbitrarily – which implicitly covers the boolean decisions from an anti-virus product.

This method starts with a window size equal to half of the file size. Both halves are occluded and the file is analyzed by the classifier. The half that results in the largest drop in classification confidence is chosen to be split for the next time step. This binary search through the file is continued until the window size is at least as small as the target window size.

The last component of our approach is to specify what method $\mathcal{D}$ should be used to replace the bytes of the original file. One approach, which we call *Random Occlusion*, is to simply select each replacement byte at random. This is similar to prior work in finding the important regions of an image according to an object detector, where it was found that randomly blocking out the pixels was effective — even if the replacement values were nonsensical [39]. We also look at *Adversarial Occlusion*, where we use a contiguous region selected randomly from one of our own benign training files to occlude the bytes in the file under test. Since we use this approach only with malicious files, Adversarial Occlusion is an especially challenging test for our machine learning based methods: they have seen these byte regions before with an explicit label of benign. This test also helps to validate that all approaches aren't simply detecting high-entropy regions that "look" packed, and then defaulting to a decision of malicious.

**Algorithm 1** Occlusion Binary Search
---
**Require:** A file $F$ of length $|F|$, a classifier $C(\cdot)$, target occlusion size $\beta$, byte replacement distribution $\mathcal{D}$
1: split $\leftarrow |F|/2$, size $\leftarrow |F|/2$
2: start $\leftarrow 0$, end $\leftarrow |F|$
3: **while** size $> \beta$ **do**
4:     $F_l \leftarrow F$, $F_r \leftarrow F$
5:     $F_l[\text{split}-\text{size}:\text{split}] \leftarrow$ contiguous sample from $\sim \mathcal{D}$
6:     $F_r[\text{split}:\text{split}+\text{size}] \leftarrow$ contiguous sample from $\sim \mathcal{D}$
7:     **if** $C(F_l) < C(F_r)$ **then**
8:         split $\leftarrow$ split $-$ size/2
9:         start $\leftarrow$ split $-$ size
10:        end $\leftarrow$ split
11:    **else**
12:        split $\leftarrow$ split $+$ size/2
13:        start $\leftarrow$ split
14:        end $\leftarrow$ split $+$ size
15:    size $\leftarrow$ size/2
16: **return** start, end
---

To show that our search procedure provides meaningful improvements for the Random and Adversarial Occlusions, we will also compare against an *Undirected Occlusion* approach that eschews our search procedure. In this case we randomly select a region of the binary to occlude with high-entropy random bytes.

Recall, that this method identifies the critical region – for a successful attack the adversary would then have to manually alter the portion of the file at that location while maintaining their intended outcome. For this reason, we use a single target window size of 2048 bytes. The resulting occluded area will fall in the range of (1024-2048] bytes. On average, this restricts the occluded section to under 0.5% of the file size for our testing set. We also limit ourselves to searching for critical regions with our n-gram model. It would be infeasible for an adversary to conduct this search across all possible classifiers, and we would expect there to be some overlap among which regions are critical. This has been shown to be true in gradient based classifiers, where an adversarial example generated by one model can fool many others [24].

### 4.3 Classification of Packed Files

Packing is a common obfuscation technique used by adversaries [22]. It also has legitimate uses, such as faster download times and protecting intellectual property. For this reason packed files cannot be assumed to be malicious. An effect of packing is that the machine instructions cannot be immediately deciphered without being unpacked first. This presents a challenge for classification using static analysis. In this experiment we use the popular UPX packer [23] to pack the benign and malicious files in our testing set. The only functional difference in the packed files is the process of unpacking itself during execution — the malicious/benign behavior of the files remains unchanged. We expect packing will greatly hinder our machine learning models, which require features to be present in the raw bytes. Packing does not completely obfuscate the file however, as data in areas like the PE header are required for a file to be executed. We observe the effect of packing on classification accuracy by rescanning the packed versions of the files.

The purpose of this test is to determine if the machine learning based models learn a degenerate "packing = malicious" decision process. This can happen easily, as the majority of packed training data came from malicious files.

### 4.4 ROP Injection

The last method we will consider is of a different variety. Instead of modifying known malicious files in an attempt to evade detection, we will inject malicious functionality into otherwise benign applications.

Many such techniques for this exist. We used the Return Oriented Programming (ROP) Injector [25] for our work. The ROP Injector converts malicious shellcode into a series of special control flow instructions that are already found inside the file. These instructions are inherently benign, as they already exist in the benign file. The technique patches the binary in order to modify the memory stack so that the targeted instructions, which are non-contiguous in the file, are executed in an order equivalent to the original shellcode. The result is that the functionality of the file is maintained, with the added malicious activity executing right before the process exits. We use the same reverse Meterpreter shellcode as Poulios et al. [25] for our experiment.

We note that not all benign executables in our testing set were injectable. The files must either have enough instructions to represent the shellcode already, or have room for further instructions to be added in a non-contiguous manner so as to prevent raising suspicion. Additionally, if the malware requires external functionality such as networking capabilities, and these are not inherent to the original executable, then portions of the PE header such as the import table must be adjusted. For these reasons, only 13,560 files were successfully infected.

Poulios et al. claim they were able to evade anti-virus over 99% of the time using this technique. This approach should be nearly undetectable with static analysis alone, as all instructions in the file remain benign in nature. This makes it an extremely difficult test for both the machine learning and anti-virus detectors.

## 5 Results

Before delving into results, we make explicit that we considered Malware the positive class and Benign the negative class, as outlined in Table 1. We remind the reader that not all tested systems are on equal footing. Our n-gram and MalConv models are trained on the same corpus, but we have no knowledge of the corpora used by the AV companies and their products. In all likelihood, they will be using datasets orders of magnitude larger than our own to ensure the quality of their products and to reach a desired false-positive rate. From this perspective alone, we would not necessarily expect the machine learning based methods to have better accuracies, as there is a disparity of information at training time. Our methods are also not optimized for the same low false-positive goal.

Table 1. True and false positive and negatives for malware classification task

|  | File is Malware | File is Benign |
| --- | --- | --- |
| Classifier says Malware | True Positive (TP) | False Positive (FP) |
| Classifier says Benign | False Negative (FN) | True Negative (TN) |

The accuracies and metrics of each method are presented in Table 2. Because of the mentioned information disparity between each system, we do not present this information for direct comparison. Our goal is to use these metrics as a baseline measure of performance for each method, and look at how these baselines are affected by various evasion techniques.

Table 2. Baseline accuracy and true/false positive/negative results for each model on the test set.

| Classifier | TN% | TP% | FN% | FP% | Accuracy% |
| --- | --- | --- | --- | --- | --- |
| N-Gram | 92.1 | 98.7 | 1.3 | 7.9 | 95.5 |
| MalConv | 90.7 | 97.2 | 2.8 | 9.3 | 94.1 |
| AV1 | 94.3 | 99.5 | 0.5 | 5.7 | 97.0 |
| AV2 | 99.4 | 64.9 | 35.1 | 0.6 | 81.6 |
| AV3 | 98.5 | 80.5 | 19.5 | 1.5 | 89.2 |
| AV4 | 93.8 | 91.9 | 8.1 | 6.6 | 92.6 |

### 5.1 Benign Modifications

The results of the experiment described in subsection 4.1 are shown in Figure 1. All four anti-virus products were significantly degraded by making seemingly insignificant changes to the malware files. The machine learning models were immune to this evasion technique. Both model's confidence that these files were malware changed by less than 1% on average. The few files that did evade were very close to the models' decision boundaries before the modifications were performed.

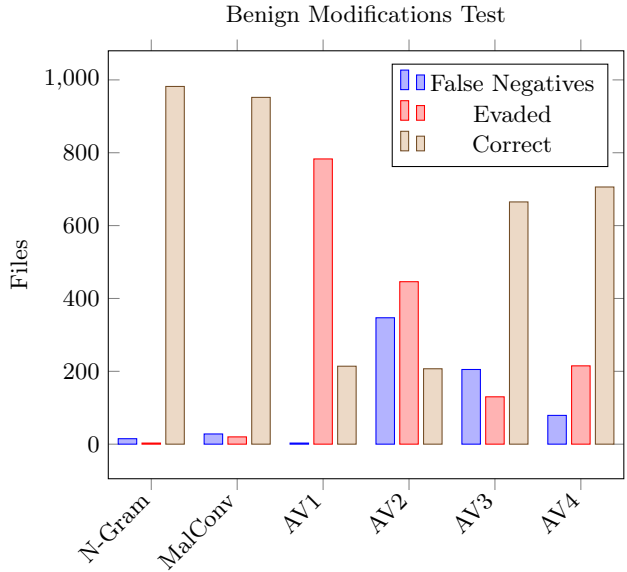

*Fig. 1.* Robustness of classifiers to benign modifications. The blue bar (left) represents the number of false negatives before any modifications were made. The red bar (center) represents the number of files that were able to evade. The tan bar (right) represents the number of files that were still classified correctly after 10 modifications. Higher is better for the tan bar, and lower is better for all others.

The anti-virus products did not perform as well in this test. AV3 and AV4 were the most robust to these modifications, with 130 and 215 files evading each respectively. While this is better than products like AV1, where 783 out of 1000 malicious files where able to evade, it is still an order of magnitude worse than the 3 and 20 files that could evade the n-gram and MalConv models, respectively.

Given these results, one might wonder — could the ML based models be evaded by this approach if more modifications were performed? In particular, a number of the modifications have multiple results for a single action. So re-applying them is not unreasonable. To answer this question, we look at the evasion rate as a function of the number of modifications performed in Figure 2. In this plot we can see that all four AV products have an upward trending slope as the number of modifications is increased. In contrast, the n-gram model is completely unchanged from 3 or more modifications, and MalConv at 4 or more. This would appear to indicate that the ML approaches are truly immune to this evasion technique, whereas the AVs are not.

These results also speak to the use of dynamic evaluation, or lack thereof, in the AV products at scan time. Any dynamic analysis based approach should be intrinsically immune to the modifications performed in this test. Because we see all AV products fail to detect 13-78% of modified malware, we can postulate that if any dynamic analysis is being done its application is ineffectual.

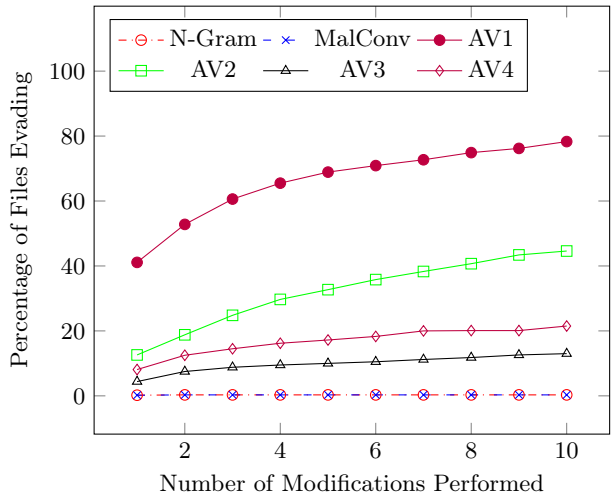

*Fig. 2.* The number of files that evade increase as the number of modifications increases. Lower is better.

We make note that in the original work of Anderson et al. [1], they applied this attack to a simple machine learning based model that used only features from the PE header. This attack was able to evade the PE-feature based model, but not the byte based models tested under our framework.

### 5.2 Targeted Occlusion

The experiment detailed in subsection 4.2 attempts to modify the most important region for detecting the maliciousness of a binary as deduced from querying only the n-gram model and attempting to transfer that knowledge to all others.

The results for the byte occlusion tests are given in Figure 3, where the blue bar shows the accuracy on 40,000 malicious files when no bytes are occluded. For both the n-gram and MalConv models, we see that the Random, Undirected, and Adversarial occlusion attacks have almost no impact on classification accuracy. In the case of the n-gram model, only 0.13% of the files where able to evade its detection after occlusion. Again evasions were highly correlated with files close to the decision boundary.

In particular, we remind the reader that the Adversarial occlusion is replacing up to 2KB of the malicious file with bytes taken from benign training data. This is designed to maximally impact the ML approaches, as the model was trained with the given bytes as having an explicit label of benign. Yet, the Adversarial choice has almost no impact on results. This is a strong indicator of the potential robustness of the ML approach, and that simply adding bytes from benign programs is not sufficient to evade their detection.

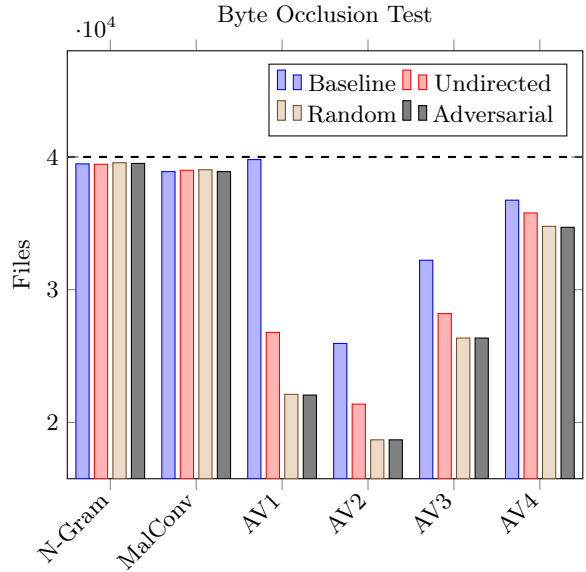

*Fig. 3.* Robustness to byte level manipulations. Black dashed line indicates maximum. The blue bar (far left) represents number of files classified correctly before occlusion. The red bar (left center) represents the number of files classified correctly after random occlusion. The tan bar (center right) represents the number of files classified correctly after targeted occlusion using random bytes. The dark gray bar (far right) represents the number of files classified correctly after targeted occlusion using bytes from known good files.

Considering the AV products we again see a different story. AV1 had the highest detection rate when no occlusion occurred, but also had the largest drop in detection rate for all three types of occlusion. AV4 had the best performance among the AV vendors, but was still impacted by the occlusion of bytes and did not perform as strongly as the ML models.

From the AV results it is also clear that the Targeted Random occlusion is an improvement over the Undirected occlusion of random bytes, showing that Algorithm 1 is effective at identifying important regions. This was not discernible from the n-gram and MalConv models, which are relatively immune to this scenario.

Although on average the occlusion was limited to 0.5% of a file's bytes, we acknowledge that one could argue this transformation may have altered the true label of the file. The search could be potentially removing the malicious payload of the application, rendering it muted (if runnable). We accept this possibility under the assumption that the adversary now has enough information to alter this small portion of the file in order to break possible signatures.

The results from the machine learning models would suggest that maliciousness is not contained to a small contiguous section of most files. This makes intuitive sense, as malicious functionality being employed in executable sections of a file might also require functions to be referenced in the import table. The non-contiguous nature of the PE file format allows for many similar circumstances, and so we believe it unlikely that all of a file's maliciousness would be contained within a small contiguous region of a binary. In addition, this results in binaries that are analogous to deployed malware with bugs. Just because malware may not function properly, and thus not negatively impact users, doesn't remove its malicious intent and the benefit in detecting it.

### 5.3 Packed Files

We observe the effect of packing on classification by comparing the baseline accuracies of the detectors to the accuracies after packing. As shown in Table 3, packing with UPX degrades the performance of all detectors, with the lone exception of AV4. The machine learning classifiers are less effective at classifying packed benign files, while the anti-virus products are worse at classifying packed malware. The disparity for the machine learning models is most likely a byproduct of unbalanced distributions of packed versus unpacked files in the benign and malicious training sets. For the anti-virus vendors, packing increases the false positive rates because the obfuscation prevents signatures from being detected. AV4's accuracy increasing is evidence that the detector has UPX unpacking capabilities built in. In terms of overall classification accuracy, AV1 suffered the largest decrease at 19%. In comparison, the n-gram and MalConv network decreased by 10% and 13% respectively.

*Table 3.* Effect of UPX Packing on Accuracy. Benign and Malware columns show the TP and TN rates on the test corpus, with the Packed columns showing the TP and TN rates after the data has been packed.

| Classifier | Benign | Packed Benign | Malware | Packed Malware |
|---|---|---|---|---|
| N-Gram  | 92.1 | 74.8 | 98.7 | 95.0 |
| MalConv | 90.7 | 68.1 | 97.2 | 92.2 |
| AV1     | 94.3 | 97.0 | 99.5 | 60.8 |
| AV2     | 99.4 | 99.3 | 64.9 | 56.5 |
| AV3     | 98.5 | 99.1 | 80.5 | 57.6 |
| AV4     | 93.8 | 93.4 | 91.9 | 95.2 |

Bulk packing remains a challenge for any static analysis classification system. The best solutions will either involve unpacking files if an unpacker exists, or relying on dynamic analysis techniques. Our results indicate that ML based systems, while also susceptible, are of comparable performance to current AV systems in this regard.

### 5.4 ROP Injection Results

In Table 4 we show the results of running the ROPInjector on the benign test files. The results are shown only for the 13,560 files that the ROPInjector was able to infect. Before infection, the Pre-ROP Accuracy column indicates the percentage of files correctly labeled benign. After infection, the Post-ROP Accuracy column indicates what percentage were correctly labeled as malicious. The Post-ROP Lift then shows how many percentage points of Post-ROP Accuracy came from the classifier correctly determining that a formerly benign file is now malicious, rather than simply being a false-positive. That is to say, if a model incorrectly called a benign file malicious, it's not impressive when it calls an infected version of the file malicious as well.

*Table 4.* Accuracy on originally benign binaries before and after applying the ROPInjector.

| Classifier | Pre-ROP Accuracy | Post-ROP Accuracy | Post-ROP Lift |
|---|---|---|---|
| N-Gram | 85.1 | 15.3 | 0.4 |
| MalConv | 82.4 | 18.8 | 1.2 |
| AV1 | 99.3 | 1.3 | 0.6 |
| AV2 | 98.7 | 1.2 | $-0.1$ |
| AV3 | 97.9 | 0.7 | $-1.4$ |
| AV4 | 89.2 | 32.9 | 22.1 |

Most malware detectors had significant difficulty with this task. The machine learning based models, n-gram and MalConv, showed only modest improvements of 0.4 and 1.2 percentage points. AV1 had a similarly small 0.6 improvement, but surprisingly AV2 and AV3 had negative improvements. That means for 1.4% of the benign files, AV3 called the original benign file malicious. But when that same file was infected by the ROPInjector, AV3 changed its decision to benign. This is a surprising failure case, and may be caused by the AV system relying too heavily on a signature based approach.

Overall these models showed no major impact from the ROPInjector. Only AV4 was able to significantly adjust its decision for 22.1% of the infected files to correctly label them as malicious. Though the evasion rate for AV4 remains high at 77%, it performed best in this scenario. Given the muted and negative performance of the other AV systems in this test, we suspect AV4 has developed techniques specifically for the ROPInjector approach.

We also notice that the benign files that were injectable include 67% of our original false positives for the n-gram model. We suspect that this is due to those files already containing the networking functionality required by the shellcode. The majority of malware requires communication over the Internet, therefore our machine learning detectors may view networking related features as somewhat malicious in nature.

Overall the ROPInjection of malicious code into otherwise benign applications presents an apparent weakness for our machine learning approaches. An important question is whether the ROPInjection is functionally invisible to the machine learning based models (i.e., it could never detect features of the injection), or is simply not sufficiently represented in our training data for the model to learn.

This questioned was tested by applying the ROPInjector to all of our training data, both benign and malicious files, which resulted in 235,865 total ROPInjected binaries. We then trained an n-gram model that tried to distinguish between ROPInjected vs Not-ROPInjected binaries. The n-gram model was able to obtain 97.6% accuracy at this task with an AUC of 99.6%. This indicates that the models could learn to detect ROPInjection, but that it is not sufficiently prevalent in our training corpus for the models to have learned.

Overall this case highlights a potential advantage of the classical AV systems with extensive domain knowledge. When sufficiently novel attacks are developed for which current models have almost no predictive power, it may be easier to develop and deploy new signatures to catch the attacks — while the machine learning approaches may require waiting for data, or creating synthetic data (which has its own risks) to adjust the model.

## 6 Discussion

We have now performed a number of tests where we modify a set of known malicious files and observe how the modifications impact ML based and classical AV systems. In performing these experiments, we note a number of interesting high level results and questions.

### 6.1 A Fairer Evaluation Scheme

In our current work the AV products have been given an advantage in potential knowledge; all the malware we are testing with has been known for some time and likely been used by the AV companies in their product updates. In an ideal test setting, all tested models would have been built with data from before a certain date X, and tested on data from files first see on on a date after X. This same kind of test was done by Saxe and Berlin [31] and found to have a significant impact on results, and more accurately reflects real use.

While such first-seen information is available from sites like VirusTotal, we unfortunately can't afford a license that will allow us to get this information for all of our data and have licensing restrictions that further complicate this. In addition, we do not have a method of obtaining outdated versions of AV products, making the evaluation further beyond our capability.

### 6.2 Detected Malware Robustness

Our initial results have shown that our Machine Learning based approaches, n-gram and MalConv, are considerably more robust to evasion techniques on malicious files that are already classified correctly. That is, once the ML based models detect a piece of malware — it appears likely the malware will require significant modifications to evade detection. This is a positive results compared to the AV systems.

Based on the Occlusion tests in particular, we can hypothesize that the ML system's robustness comes from using features dispersed through the whole of the binary. If only a small portion of the file was used to make its conclusion, the Adversarial Occlusion test in particular would have been more effective. In contrast, the way signatures are designed encourages specificity. Using markers from broad portions of the file to create a signature increases the risk that false positives occur, which AV vendors are often trying to minimize.

### 6.3 Generic Adversarial Approaches to Malware

This work was done for Windows x86-x64 based binaries. But malware is a problem that enters every new computing environment. One question worth asking is, do the benefits (or costs) of a ML based system transfer to these other domains as well? The byte occlusion test we developed in subsection 4.2 works without knowledge of its domain, and so could be re-used for other file formats (such as PDFs) or operating systems (such as Linux or iOS). Beyond simply developing new tests for each domain format, what other tests can be developed which can easily cross domains?

### 6.4 Executable Data Augmentation

We also take a moment to note an unexploited similarity between our work and prior results in image classification. Augmenting the training data through the use of content-preserving distortions (rotations, color alterations, cropping, etc.) has long been used for training neural networks [18]. One question is whether the evasion techniques used in this work could instead be used for improving the machine learning models through a similar augmentation scheme. The results with ROP injection in subsection 5.4 highlight a case where this could help the models to learn to catch an attack they are currently missing. However, if any of the techniques are used to perform augmented training, it may then become invalid to use the same techniques to quantify the benefit of a Machine Learning approach.

For example, training on ROPInjected binaries may only improve the model's ability to detect that specific type of adversarial alteration. For malware this is a particularly important distinction, as the nearly-unlimited degrees of freedom that are afforded by the Windows PE format may allow many new and novel counter-measures, and our goal is for machine learning based systems to generalize to *new* types of malware, never before seen. Thus training on the ROPInjector may improve the apparent accuracy on *current* malware, but not improve its performance with respect to *novel* malware. Measuring and quantifying this concern is an important area for future work.

# 7 Conclusion

We have demonstrated a new testing methodology for comparing the robustness of machine learning classifiers to current anti-virus software. Furthermore, we have provided evidence that machine learning approaches may be more successful at catching malware that has been manipulated in an attempt to evade detection. Anti-virus products do a good job of catching known and static malicious files, but their rigid decision boundaries prevent them from generalizing to new threats or catching evolutionary malware. We demonstrated that top tier anti-virus products can be fooled by simple modifications including changing a few bytes or importing random functions. The machine learning models appear to better generalize maliciousness — leading to an increased robustness to evasion techniques compared to their anti-virus counterparts. Packed files and other obfuscation techniques remain a challenge for pure static analysis systems, but future work can be done to incorporate the power of machine learning for dynamic analysis as well.